\documentclass[aps,prl,twocolumn,epsfig,preprintnumbers,superscriptaddress,10pt]{revtex4}

\usepackage{graphicx}
\usepackage{dcolumn}
\usepackage{bm}

\newcommand{\be}{\begin{equation}}
\newcommand{\ee}{\end{equation}}
\newcommand{\bea}{\begin{eqnarray}}
\newcommand{\eea}{\end{eqnarray}}
\newcommand{\ba}{\begin{eqnarray}}
\newcommand{\ea}{\end{eqnarray}}

\newcommand{\beq}{\begin{equation}}
\newcommand{\eeq}{\end{equation}}
\newcommand{\beqa}{\begin{eqnarray}}
\newcommand{\eeqa}{\end{eqnarray}}
\newcommand{\beqar}{\begin{eqnarray*}}
\newcommand{\eeqar}{\end{eqnarray*}}

\def\co{{\mathcal C}_{\rm vac}}
\def\ct{{\mathcal C}_{\rm targ}}

\begin{document}

\preprint{CERN-PH-TH/2011-233}

\title{Medium-induced color flow softens hadronization}

\author{Andrea Beraudo}
\affiliation{Physics Department, 
    Theory Unit, CERN,
    CH-1211 Gen\`eve 23, Switzerland}
\affiliation{Centro Studi e Ricerche "Enrico Fermi", Piazza del Viminale 1,  Roma, Italy}    
\author{Jos\'e Guilherme Milhano}
\affiliation{Physics Department, 
    Theory Unit, CERN,
    CH-1211 Gen\`eve 23, Switzerland}
\affiliation{CENTRA, Instituto Superior T\'ecnico, Universidade T\'ecnica de Lisboa,
Av. Rovisco Pais, P-1049-001 Lisboa, Portugal}
\author{Urs Achim Wiedemann}
\affiliation{Physics Department, 
    Theory Unit, CERN,
    CH-1211 Gen\`eve 23, Switzerland}


\begin{abstract}
Medium-induced parton energy loss, resulting from gluon exchanges between the QCD
matter and partonic projectiles, is expected to underly the strong suppression of
jets and high-$p_T$ hadron spectra observed in ultra-relativistic heavy ion collisions. Here, we
present the first color-differential calculation of parton energy loss. We find that color
exchange between medium and projectile enhances the invariant mass of energetic color singlet 
clusters in the parton shower by a parametrically large factor proportional to the 
square root of the projectile energy. This effect is seen in more than half of the most energetic 
color-singlet fragments of medium-modified parton branchings. Applying a standard cluster
hadronization model, we find that it leads to a characteristic additional 
softening of hadronic spectra. A fair description of the nuclear modification factor measured at the 
LHC may then be obtained for relatively low momentum transfers from the medium. 
\end{abstract}

\maketitle


High transverse momentum partons ($p_T \gg 10$ GeV) produced in heavy ion collisions
interact with the QCD matter in the collision region while branching. This interaction is thought 
to cause the strong medium modification of hadronic spectra and jets measured in heavy ion 
collisions at the LHC and at RHIC. The modeling of this jet quenching 
phenomenon has focussed 
so far on medium-induced parton energy loss prior to hadronization~\cite{Armesto:2011ht}, 
assuming that for sufficiently high $p_T$ hadronization occurs time-delayed  
outside the medium. However, if the color flow 
of a parton shower is modified within the medium, then hadronization can be affected  irrespective of
when it occurs.
Here, we analyze for the first time the medium-induced color flow in a standard
parton energy loss calculation.  Compared to the current modeling of parton energy loss~\cite{Chen:2011vt,Arleo:2011rd,Horowitz:2011gd,Lokhtin:2011qq,Zakharov:2011dq,Fochler:2011en},
this will be seen to result in a characteristic softening of
the ensuing hadronization. It may thus affect significantly the extraction of medium properties
from the measured nuclear modification factor at the LHC~\cite{QM11ALICE,QM11CMS,Aamodt:2010jd}.

We start by considering an elementary building block of a parton shower, the 
$q \to q\, g$ parton splitting. For a small light-cone energy $k^+$ of the gluon compared to the parent
parton, $x \equiv k^+/p^+ \ll 1$, and for transverse gluon momentum ${\bf k}$ with ${\bf K_0} \equiv  {\bf k}/ {\bf k}^2$,
the gluon spectrum reads, to leading order in  $\alpha_s$,
\begin{equation}
k^+\frac{dI^{\rm vac} }{dk^+\, d{\bf k}}=\frac{\alpha_s C_R}{\pi^2}\,{\bf K_0}^2\, .
\label{eq1}
\end{equation}
In the presence of QCD matter, interactions between projectile parton and medium result in 
a modified spectrum $dI^{\rm rad}\!\equiv\!dI^{\rm vac}\!+\!dI^{\rm med}$ that can be written as 
an expansion in powers of the ratio $\zeta\equiv L^+/\lambda^+_{\rm el}$ of in-medium 
path length $L^+$ and an elastic mean free path $\lambda^+_{\rm el}$
(for details, see~\cite{Armesto:2011ht}).
To first order in this opacity expansion, the medium-induced radiation spectrum reads
\begin{equation}
k^+\frac{dI^{\rm med}}{dk^+\, d{\bf k}}\! =\!\zeta \frac{\alpha_s C_R}{\pi^2}\!
	\bigg\langle\! \Big(\! 
	({\bf K_0}-{\bf K_1})^2\!-\!{\bf K_0}\!^2  + {\bf K_1}\!^2 
	\Big) \!{\cal T_I} \!\bigg\rangle .
		\label{eq2}
\end{equation}
%
%
Here, $\langle \dots \rangle$ denotes averaging over the transverse momentum transfer ${\bf q}$ of a single
interaction, and ${\bf K_1} \equiv ({\bf k}- {\bf q})/ ({\bf k}- {\bf q})^2$. 
Medium-induced interference effects enter via the factor 
\begin{equation}
 {\cal T_I} = \left( 1 - \frac{\sin \left(\omega^-_1\, L^+ \right) }{\omega^-_1\, L^+} \right) =
          \left\{ \begin{array} 
           {r@{\quad  \hbox{for}\quad}l}
             1 & 1/\omega^-_1 \ll L^+ \, ,\\
             0 & 1/\omega^-_1 \gg L^+\, .
          \end{array} \right.
        	\label{eq3}
\end{equation} 
Thus,  the medium-modification $dI^{\rm med}$ 
can occur only for quanta of sufficiently
small formation time $1/\omega^-_1 \equiv 2k^+/\left({\bf k}-{\bf q} \right)^2 \ll L^+$. 
%
\begin{figure}[!tp]
\begin{center}
\includegraphics[clip,width=.48\textwidth]{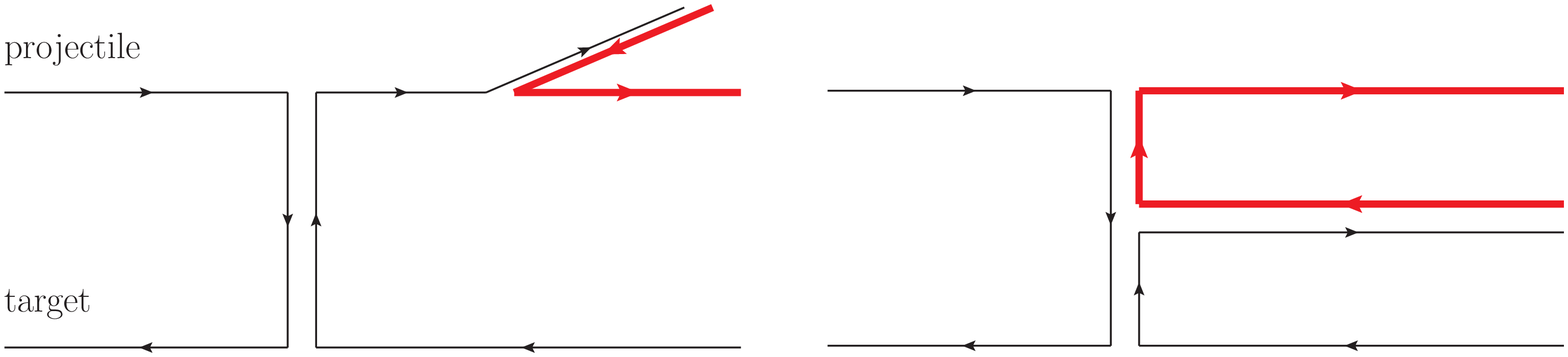}
\includegraphics[clip,width=.48\textwidth]{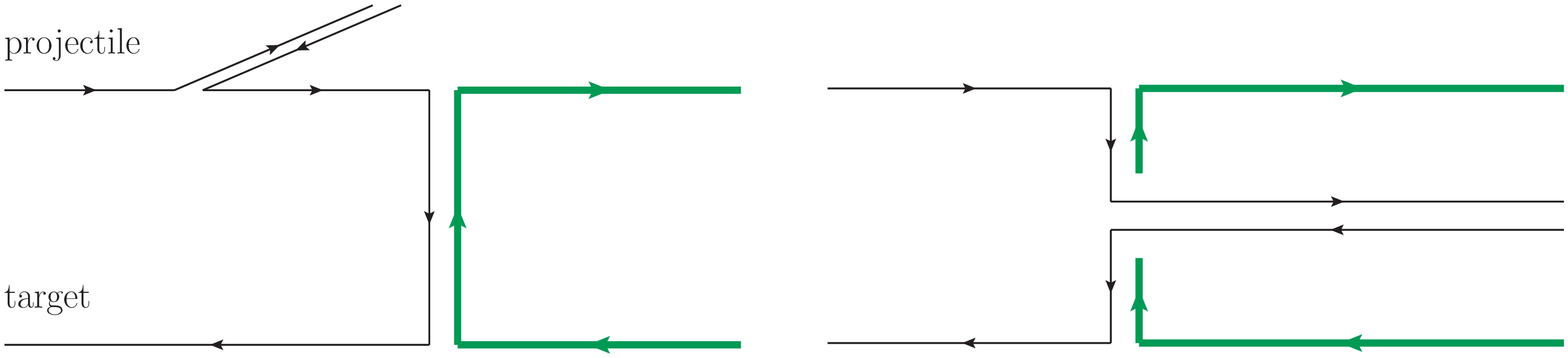}
\caption{$N=1$ opacity diagrams for gluon radiation from a projectile quark
in the large-$N_c$ limit. The most energetic color-singlet clusters are denoted
by thicker lines and correspond to color flow between projectile components
(upper diagrams, contribution $dI^{\rm med}_{{\rm low}\, M}$) or between projectile 
and target  (lower diagrams, contribution $dI^{\rm med}_{{\rm high}\, M}$). Diagrams on the right hand
side include a 3-gluon vertex.}
\label{fig1}
\end{center}
\vspace*{-.4cm}
\end{figure}
%
To first order
in opacity and at large $N_c$, we identify three contributions with distinct color flow
\begin{equation}
{dI^{\rm med}}={dI^{\rm med}_{\rm virt}}+{dI^{\rm med}_{{\rm low}\, M}}+{dI^{\rm med}_{{\rm high}\, M}}\,.
\label{eq4}
\end{equation}
Here, $dI^{\rm med}_{\rm virt}$ arises from probability-conserving 
{\it virtual} corrections that do change neither color flow nor
kinematic distributions in the projectile. The contributions
$dI^{\rm med}_{{\rm low}\, M}$ and $dI^{\rm med}_{{\rm high}\, M}$ contain medium effects and 
can be characterized by the invariant mass of their leading cluster, as we discuss now.

For a projectile with (light-cone) energy much larger than that of its target scattering partner,
$p^+\gg t^+$, and for $x\ll 1$, the most energetic parton in the final state in Fig.~\ref{fig1} is defined 
unambiguously and carries energy  $p_f^+\equiv (1-x)\, p^+$. Distributing the energy of the gluon 
equally between its $q$ and $\bar{q}$-legs in the large $N_c$-limit (our final conclusions will not
depend on this assumption), 
the energy and invariant mass of the most energetic singlet cluster in a vacuum splitting read
\begin{equation}
	P^{+}_{\co} \simeq \left( 1 + x/2 \right) p_f^+ 
	\, , \quad M^2_{\co} 
	\simeq {\bf k}^2/ 2\, x\, .
	\label{eq5}
\end{equation}
These relations hold for the leading color singlet clusters of the vacuum contribution (\ref{eq1}), 
for the virtual correction $dI_{\rm virt}^{\rm med}$ in (\ref{eq4}), and for 
$dI^{\rm med}_{{\rm low}\, M}$ (in the latter case,  the distributions in $x$ and ${\bf k}$ contain 
information about the medium modification). In contrast, for $dI^{\rm med}_{{\rm high}\, M}$ 
color flows from the most energetic final state parton directly to a target component of
momentum $t$, and
\begin{equation}
	P^{+}_{\ct} \simeq p_f^+ 
	\, , \quad M^2_{\ct}
	= (p_f+t)^2 	
	\simeq  p^+\, Q_T\, .
	\label{eq6}
\end{equation}
Here, $Q_T \equiv \sqrt{2} E_{\rm th}$ and $E_{\rm th}$ is the typical (thermal) energy of the target component. 
The invariant mass in (\ref{eq6}) is parametrically larger than in the vacuum case (\ref{eq5}) by a factor $\sqrt{p^+}$. 
This motivates the choice of subscripts in $dI^{\rm med}_{{\rm low}\, M}$ and $dI^{\rm med}_{{\rm high}\, M}$. 

One can determine the kinematic conditions and time scales required for $dI^{\rm med}_{{\rm high}\, M}$ 
to contribute. The color-inclusive sum (\ref{eq4}) depends on the formation time $1/\omega_1^-$. 
To first order in opacity, 
we find that the color-differential contributions on the right hand side of~(\ref{eq4}) 
carry interference factors involving two formation times $1/\omega_1^-$ and 
$1/\omega_0^- \equiv 2k^+/{\bf k}^2$ (see ~\cite{bmw} for full result). Here, we focus on the relevant limiting cases.

If parton energy loss becomes negligible ($\omega_1^-\, L^+ \ll 1$) 
and if the formation time of the final state gluon is large ($\omega_0^-\, L^+ \ll 1$), then
\begin{eqnarray}
	k^+ \frac{dI^{\rm med}_{\rm virt}}{dk^+\, d{\bf k}} &=& 
	- \zeta\,
	 \frac{\alpha_s C_R}{2\pi^2} \big\langle  {\bf K_0}^2 \big\rangle\, ,
	\nonumber \\
	k^+ \frac{dI^{\rm med}_{{\rm low}\, M}}{dk^+\, d{\bf k}} &=& 
	 \zeta\,
	 \frac{\alpha_s C_R}{2\pi^2} \big\langle  {\bf K_0}^2 \big\rangle\, ,
	\quad 
	k^+ \frac{dI^{\rm med}_{{\rm high}\, M}}{dk^+\, d{\bf k}} = 0\, .
	\label{eq7}
\end{eqnarray}
In this case, the color exchange between the medium and the projectile
occurs predominantly at early times {\it before} the splitting and hence the color flow and invariant
mass of the leading color singlet is vacuum-like, $dI^{\rm med}_{{\rm high}\, M} = 0$. 
In contrast, in the limit $1/\omega_0^- \ll L^+ \ll  1/\omega_1^-$, one finds that 
$dI^{\rm med}_{\rm virt}\sim -3\langle  {\bf K_0}^2 \rangle $,  
$dI^{\rm med}_{{\rm low}\, M}\sim \langle  {\bf K_0}^2 \rangle $, and  
$dI^{\rm med}_{{\rm high}\, M}\sim 2 \langle  {\bf K_0}^2 \rangle $. 
Thus, even if the color-inclusive sum (\ref{eq4}) vanishes,  
there is a kinematic regime in which medium-induced color flow changes the invariant mass of the 
leading color singlet significantly.
 
If there is parton energy loss, one finds in the limit
$1/\omega_1^-\, ,  1/\omega_0^- \ll L^+$
{\setlength\arraycolsep{1pt}
\begin{eqnarray}
	k^+ \frac{dI^{\rm med}_{\rm virt}}{dk^+\, d{\bf k}} &=& 
	- 3\, \zeta\, \frac{\alpha_s C_R}{2\pi^2}
	\big\langle  {\bf K_0}^2 \big\rangle\, ,
	\nonumber \\
	k^+ \frac{dI^{\rm med}_{{\rm low}\, M}}{dk^+\, d{\bf k}} &=& 
	 \zeta\, \frac{\alpha_s C_R}{2\pi^2}
	\big\langle ({\bf K_0}-{\bf K_1})^2+ {\bf K_1}^2  \big\rangle\, ,
	\nonumber \\
	k^+ \frac{dI^{\rm med}_{{\rm high}\, M}}{dk^+\, d{\bf k}} &=& 
	 \zeta\, \frac{\alpha_s C_R}{2\pi^2}
	 \big\langle  ({\bf K_0}\!-\!{\bf K_1})^2\!+\! {\bf K_1}^2 \!+\! {\bf K_0}^2 
	  \big\rangle.		
	  \label{eq8}
\end{eqnarray}}
In summary, whenever the color-averaged contribution (\ref{eq4}) to radiative energy loss is
significant, the resulting medium modification of color flow changes the kinematic  properties
of the most energetic color singlet in more than 50\% of all cases, 
$dI^{\rm med}_{{\rm high}\, M} > 0.5\,  dI^{\rm med}$.
%

For a gluonic projectile, one finds~\cite{bmw} 
to first order in opacity, and in the large $N_c$ limit a similar fraction of clusters $\ct$
of parametrically large invariant mass (\ref{eq6}). To higher orders in opacity, multiple 
interactions between projectile and target enhance the fraction of color singlet clusters $\ct$~\cite{bmw}.  
Therefore, the calculation above illustrates generic features of medium-modified color flow in
models of parton energy loss. 

Since color is conserved in QCD and since hadrons are color singlets, a dynamically consistent
hadronization prescription must relate hadrons to color singlet fragments of the parton shower. We
therefore expect on general kinematic grounds that the medium-modified distribution (\ref{eq8})
of leading color singlets has consequences for the dynamics of hadronization. We now illustrate this
point in a cluster hadronization model that encodes essential features of 
the prescription implemented in the MC event generator HERWIG~\cite{HERWIG}; 
 the parton shower is evolved perturbatively to a hadronic scale 
($Q_0 \simeq 600$ MeV in Ref.~\cite{HERWIG})
at which the entire shower is decomposed into its color singlet components (`clusters'). Clusters $\cal C$
of low invariant mass (${\cal M}_{\cal C} <   {\cal M}_{\rm cr} = 4 \, {\rm GeV}$) 
are then decayed directly into pairs of hadrons,
while clusters of larger invariant mass are decayed first into pairs of daughter clusters,
${\cal C} \to X\, Y$, till the invariant mass of the daughters satisfies the condition for decay into hadrons.
In parton showers evolved in the vacuum, far more than 90\% of all clusters are found to have
a small invariant mass, ${\cal M}_{\cal C} < 4$ GeV~\cite{HERWIG}. We therefore take in the following
${\cal M}_{\cal C_{\rm vac}} < 4\,{\rm GeV}$. We want to adapt this cluster 
hadronization model to a medium-modified parton shower, for which at the final scale $Q_0$
the most energetic parton carries energy $p_f^+$.  
In more than 50\% of these showers, color will flow from this parton 
directly to the target, and one finds for the most energetic cluster $\ct$ an
invariant mass ${\cal M}_{\ct} > 4\,{\rm GeV}$ within the 
range $p_f^+>p_{\rm cr}^+={\cal M}_{\rm cr}^2/Q_T$.
Estimating the thermal energies of target scatterers by their ideal gas value $E_{\rm th} \simeq 2.7\, T$,
we find for $T = 200$ (500) MeV  a value of $Q_T \simeq 760$ (1900) MeV corresponding to
$p_{\rm cr}^+ = 21.0 (8.4)$ GeV. Therefore, for a wide, phenomenologically relevant range of parton
energies
$p_f^+>p_{\rm cr}^+$, hadronization of $\ct$ involves an additional step $\ct \to X\, Y$
that is absent in the fragmentation of cluster ${\cal C}_{\rm vac}$, and that leads
to a most energetic daughter cluster $X$ of energy~\cite{HERWIG}
\begin{equation}
	P_X^+ = \left(1 - \frac{Q_0}{{\cal M}_{\ct}} \right)\, p_f^+ + \frac{Q_0}{{\cal M}_{\ct}}
	   t^+\, .
	   \label{eq9}
\end{equation}
Therefore, amongst the distribution of {\it final} clusters ${\cal C}_f$, defined as
those that decay directly into hadrons, the daughters
of $\ct$ are significantly softer (${\cal C}_f=X$, $P_X^+ < p_f^+$) than ${\cal C}_{\rm vac}$
(${\cal C}_f={\cal C}_{\rm vac}$, $P_{{\cal C}_{\rm vac}}^+ \geq p_f^+$), see (\ref{eq5}).

\begin{figure}[!tp]
\begin{center}
\includegraphics[clip,width=.45\textwidth]{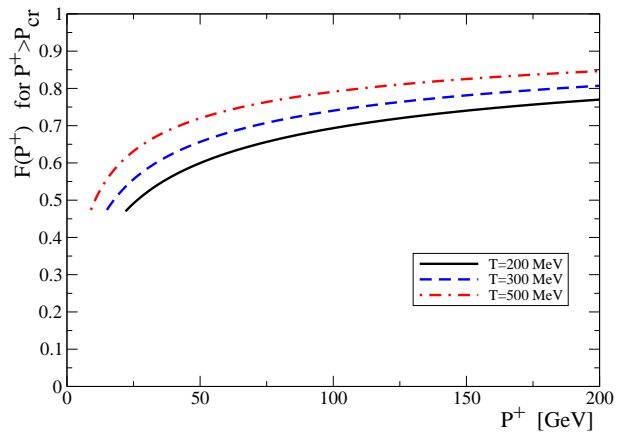}
\caption{The  color-induced suppression factor $F(P^+)$ of (\ref{eq12}) plotted for sufficiently large
cluster energies $P^+ > P_{\rm cr}$ where additional color-flow induced 
cluster decays occur. 
}
\label{fig2}
\end{center}
\vspace*{-.4cm}
\end{figure}

To illustrate how the softening of the cluster distribution due to color flow may affect transverse momentum
spectra, we consider a sample of hadronic collisions with final state parton showers evolved to scale 
$Q_0$. We parametrize the resulting transverse partonic spectrum by a power-law
(within the geometry of a heavy ion collision, the large momentum component of a hard process
points transverse to the beam direction) 
\begin{equation}
	\frac{dN}{d{p_f^+}} = \frac{c}{{p_f^+}^n}\, .
	\label{eq10}
\end{equation}
A fraction $f_t$ of the partons in (\ref{eq10}) are endpoints of clusters ${\cal C}_{\rm targ}$
with color flowing directly to the target. The distribution of final clusters ${\cal C}_f$ resulting
from $\ct$ reads
\begin{eqnarray}
	&&\!\!\!\!\!\!
	\frac{dN^{{\cal C}_f}}{d{P^+}} = f_t \int_0^{P_{\rm cr}^+} dp_f^+\, 
		\delta\left(P^+-p_f^+ \right)\, \frac{dN}{d{p_f^+}} \nonumber\\
		&&\quad + f_t \int_{P_{\rm cr}^+}^{\infty} dp_f^+\, 
		\delta\left(P^+-p_f^+ \left(1 - \frac{Q_0}{\sqrt{p_f^+\, Q_T}} \right)  \right)\, \frac{dN}{d{p_f^+}} \nonumber\\
		&& \qquad = f_t \frac{c}{{P^+}^n}  
				F(P^+)\, , \label{eq11}\\
%
%
	&&\!\!\!\!\!\!\!\!\!F(P^+) = \Theta(P^+_{\rm cr} - P^+) \nonumber \\
	&&\quad \quad + \Theta(P^+ - (P^+_{\rm cr}-\Delta) )
	 \frac{2\, \left(1+\frac{Q_0}{\sqrt{P^+\, Q_T}}\right)^{1-n}}{\left(2+\frac{Q_0}{\sqrt{P^+\, Q_T}}\right)} 	 \, .
	\label{eq12}
\end{eqnarray}
Here, $\Delta \equiv Q_0 \sqrt{P_{\rm cr}^+/Q_T}$,
and the last factor in (\ref{eq12}) comes from transforming (\ref{eq10}) from $p_f^+$ to 
$P_X^+$ with the help of (\ref{eq9}). The first term in (\ref{eq12}) 
comes from clusters ${\cal C}_{\rm targ}$ of small energy $P^+ < P^+_{\rm cr}$
and small invariant mass,  that decay directly into hadrons, ${\cal C}_f = {\cal C}_{\rm targ} $. The
second term comes from clusters ${\cal C}_{\rm targ}$ of large invariant mass that undergo a cluster
decay, ${\cal C}_f = X$. Both mechanisms contribute in an intermediate range 
$P^+_{\rm cr}-\Delta < P^+< P^+_{\rm cr}$ where one finds an enhancement $F(P^+)> 1$. 
A more realistic treatment, including corrections to  the limit 
$p_f^+ \gg t^+$, may be expected to smoothen the $\Theta$-functions in (\ref{eq12}). Here, we steer
clear of these model-dependent uncertainties by focussing on the 
high-energy region $P^+> P^+_{\rm cr}$ where an additional cluster decay results in a
suppression, $F(P^+)<1$.  For phenomenologically motivated parameter choices, namely 
a power-law spectrum (\ref{eq10}) with $n=6$, values $Q_0 = 600$ MeV
and ${\cal M}_{\rm cr} = 4$ GeV typical of hadronization models, and a range of expected
thermal energies, the color-induced suppression persists up to cluster energies well above 
100 GeV, see Fig.~\ref{fig2}.

Medium-effects in single inclusive hadron spectra are typically expressed
in terms of the nuclear modification 
factor $R_{\rm AA}(p_T) \equiv \left(dN^{\rm AA}/dp_T\right)/ 
\left(n_{\rm coll}\, dN^{\rm pp}/dp_T\right)$. Let us denote by $R_{\rm AA}^{\rm fact}$
the nuclear modification factor  calculated in a standard implementation of parton energy loss 
in which the quenching dynamics is factorized from hadronization and the latter is treated as
in the vacuum~\cite{Armesto:2011ht}. In this case, the energy degradation due to (\ref{eq4}) is accounted for in
$R_{\rm AA}^{\rm fact}$, whereas effects of medium-induced color flow are not. 
The latter effects may be included by subjecting a fraction $f_t$ of the hadronic yield 
to a further suppression (\ref{eq12}) at the cluster level. We do this by 
relating the energy of the final clusters ${\cal C}_f$  to the hadronic transverse momentum, 
$P^+ = \sqrt{2} \frac{p_T}{z}$, and approximating the momentum fraction $z$ carried by the hadron
by the average of a boosted, isotropic two-body decay, $\langle z\rangle \sim 3/4$,
\begin{eqnarray}
	R_{\rm AA}(p_T) &\simeq& \left(1-f_t\right)\, R_{\rm AA}^{\rm fact}(p_T)
		\nonumber \\
	                              && + f_t\, F\left(\sqrt{2}\frac{4}{3}p_T\right)\, R_{\rm AA}^{\rm fact}(p_T)\, .
	                                  \label{eq13}
\end{eqnarray}
%
\begin{figure}[!tp]
\begin{center}
\includegraphics[clip,width=.45\textwidth]{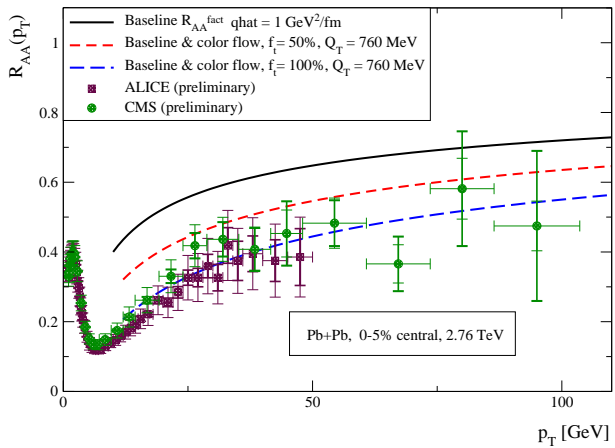}
\caption{The nuclear modification factor $R_{\rm AA}(p_T)$. The baseline calculation of kinematic
effects (solid black curve) is supplemented with the effect of color-flow 
modified hadronization according to (\ref{eq13}). }
\label{fig3}
\end{center}
\vspace*{-.4cm}
\end{figure}
%
The resulting $R_{\rm AA}$ is plotted
in Fig.~\ref{fig3} starting from a standard implementation of parton energy loss
for central Pb-Pb collisions at the LHC (curve for $R_{\rm AA}^{\rm fact}$  
taken from Fig.~5 of Ref.~\cite{Eskola:2004cr}). We chose
this baseline since - in contrast with many results in the recent literature - it refers to
a relatively small density of the medium (characterized by the quenching parameter 
$\hat{q}= 1\, {\rm GeV}^2/{\rm fm}$), and therefore significantly overpredicts the $R_{\rm AA}$ measured 
at the LHC. Supplementing this model with a color-flow induced suppression $F$ in $f_t = 50 \%$  (100\%) 
of all fragmentation patterns according to (\ref{eq13}), we find that $R_{\rm AA}$ is reduced significantly 
by $\sim 0.1$ ($\sim 0.2$) compared to $R_{\rm AA}^{\rm fact}$. For $f_t = 100 \%$, 
this brings the model into reasonable agreement 
~\footnote{In~\cite{Eskola:2004cr}, results for $R_{\rm AA}^{\rm fact}$ at $\sqrt{s_{\rm NN}}= 5.5$ TeV
and $\sqrt{s_{\rm NN}}= 2.76$ TeV differ by much less than the curves shown for different
the color-flow induced suppression in Fig.~\ref{fig3}. Therefore, although calculated for 
$\sqrt{s_{\rm NN}}= 5.5$ TeV, we regard the black straight line in Fig.~\ref{fig3} as suitable
for a semi-quantitative comparison with data.}
with preliminary results of ALICE and CMS at the LHC.

Recent measurements of the nuclear modification factor $R_{\rm AA}$ in central Pb-Pb collisions 
at the LHC have established unambiguously that $R_{\rm AA}(p_T)$ 
increases notably at intermediate transverse momentum  $p_T\in [5; 20]$ GeV, 
and they provided first information 
about the continuing slow rise of $R_{\rm AA}(p_T)$ 
in the range $p_T \in [20;100]$ GeV. This has been used
in several recent studies~\cite{Chen:2011vt,Arleo:2011rd,Horowitz:2011gd,Lokhtin:2011qq,Zakharov:2011dq,Fochler:2011en} to constrain medium properties entering the modeling of parton energy loss. 
It is a common feature of all these models to
assume gluon exchanges between the projectile parton and the medium,
while neglecting in their dynamical implementation the ensuing changes in the color structure of the parton 
shower.  Although heuristic models of medium-modified hadronization have been discussed
previously~\cite{Sapeta:2007ad,Zapp:2008gi}, we assessed here 
for the first time the color-differential information that is implicitly contained in standard
implementations of parton energy loss. 
From Figs.~\ref{fig2} and ~\ref{fig3} we learn that medium-modified color flow may
affect single inclusive hadron spectra significantly up to the highest transverse momenta 
($p_T > 100$ GeV, say). Also, remarkably, in contrast with the temperature dependence of the
baseline $R_{\rm AA}^{\rm fact}$, 
the smaller the thermal energy in (\ref{eq12}), the larger the color-flow induced contribution to the
suppression of $R_{\rm AA}$. 
As a consequence, a significantly smaller density of the medium may then be 
sufficient to account for the observed suppression of $R_{\rm AA}$. 

Medium-modified color flow is an inevitable
consequence of models of parton energy loss. 
However, its numerical manifestation in $R_{\rm AA}$ may depend sensitively on the microscopic 
implementation of parton energy loss, including e.g. the spatio-temporal embedding of the parton shower
 in the QCD matter and the probability that the produced partons escape the medium without gluon exchange.
As illustrated in Fig.~\ref{fig3}, the effect of medium-induced color flow is potentially large
and thus needs to be constrained in phenomenological applications. 
It is thus important to include the effect of  medium-modified color flow 
in full microscopic simulations of parton energy loss,  where also refined or different 
hadronization prescriptions 
can be explored.  Such further studies could give insight into the existence or
absence of changes in the hadrochemistry and in the fragmentation pattern of jets in heavy 
ion collision. They could also help to understand any possible difference between the $R_{\rm AA}$ of
reconstructed jets and leading hadrons. 

We thank G. Corcella and K. Zapp for helpful discussions.
JGM acknowledges the support of Funda\c c\~ao para a Ci\^encia e a Tecnologia (Portugal) under project CERN/FP/116379/2010.

\end{document}